\documentclass[sigconf]{acmart}

\usepackage{graphicx}
\usepackage{subcaption}

\AtBeginDocument{%
  \providecommand\BibTeX{{%
    \normalfont B\kern-0.5em{\scshape i\kern-0.25em b}\kern-0.8em\TeX}}}

\setcopyright{acmcopyright}
\copyrightyear{2020}
\acmYear{2020}

\acmConference[KDD Humanitarian Mapping Workshop '20]{ACM SIGKDD Conference on Knowledge Discovery and Data Mining
}{August 24, 2020}{San Diego, California USA}
\acmBooktitle{KDD Humanitarian Mapping Wokshop '20: ACM SIGKDD Conference on Knowledge Discovery and Data Mining
August 24, 2020; San Diego, California USA}

\begin{document}

\title[Mapping potential informal settlements in Tegucigalpa with ML]{Case study: Mapping potential informal settlements areas in Tegucigalpa with machine learning to plan ground survey}


\author{Federico Bayle}
\affiliation{%
  \institution{Dymaxion Labs}
  \city{Buenos Aires}
  \country{Argentina}}
\email{federico@dymaxionlabs.com}

\author{Damian E. Silvani}
\affiliation{%
  \institution{Dymaxion Labs}
  \city{Buenos Aires}
  \country{Argentina}}
\email{damian@dymaxionlabs.com}

\renewcommand{\shortauthors}{Bayle and Silvani}

\begin{abstract}

Data collection through censuses is conducted every 10 years on average in Latin America, making difficult monitoring the growth and support needed by communities living in informal settlements. Conducting a field survey requires logistical resources to be able to do it exhaustively. The increasing availability of spatial open-data, high-resolution satellite images, and open-source tools, allow us to train machine learning algorithms map these areas in different cities of Latin America. This case study shows the collaboration between Dymaxion Labs and the NGO Techo to employ that techniques to create the first informal settlements census of Tegucigalpa, Honduras.

\end{abstract}





\acmISBN{}
\acmDOI{}

\maketitle

\section{Introduction}

Poverty is a major issue in Latin America. Slums and informal settlements grow quickly, and there is a need for more up-to-date data for policymaker decisions. If governments had updated information about slums and their growth, they could give the affected families better life conditions and change their future. Having updated information helps to improve health, education, and security for the children who currently live there. In order to carry out an accurate survey of informal areas, exhaustive coverage of the territory is required. This task requires having costly logistical and material resources, which in turn undermines periodicity and scope.

These settlements often do not have critical public services such as sanitation, resulting in health and environmental hazards, especially for children. 

Tegucigalpa (formally Tegucigalpa, Municipality of the Central District) is Honduras' largest and most populous city as well as the nation's political and administrative center. The urban population is estimated to 1,143,373 inhabitants by 2019. The urban area is 201.5 square kilometers. There no significant progress has been seen  in this matter in recent decades about informal settlements life conditions.  The data required to formulate public policies aimed at improving the situation of informal settlements population, is almost non-existent through official and unofficial sources in Honduras. 

For this reason, the NGO TECHO\footnote{\url{www.techo.org}} planned to conduct in 2018 a census of informal settlements in the urban district of the Central District, aiming to provide clear, reliable and public information on the current problem of thousands of people living in poverty. TECHO is a youth-led non-profit organization founded in 1997 with a presence in Latin America and the Caribbean. They seek to build a fair, integrated, and poverty-free society, where everyone has the opportunities needed to develop their capacities and fully exercise their rights.

Due to the lack of the previous census and public data about informal settlements in Tegucigalpa, in Dymaxion Labs\footnote{\url{www.dymaxionlabs.com}} we partnered with TECHO to help them map the potential areas to survey. The main objective was to reduce the survey area to be scouted by the volunteer team. Based on our previous work on informal settlements mapping in Buenos Aires (Argentina) and Asunción (Paraguay), the machine learning algorithm AP-LATAM\cite{aplatamweb} looked promising to apply in Honduras based on a previous survey of Guatemala City (Guatemala). Dymaxion Labs is a startup developing a cloud-based API to detect objects in satellite imagery at scale.

\section{Related Work}
Following Kuffer, Pfeffer, and Sliuzas (2016)\cite{Kuffer2016SlumsFS}, at least 87 papers were published in English about this topic. 
Reviewing some of them, we can find studies for Argentina, Colombia, England, Ghana, Nigeria, Tanzania, Uganda, Malawi, and Rwanda. 

Focusing on Latin American cities, Hall, Malcolm, and Piwowar (2001)\cite{hall2001integration} used Landsat and Radarsat imagery combined with GIS data to detect urban poverty pockets in Rosario, Argentina. Patiño and Duque (2013)\cite{Patino2013ARO} used very high-resolution images to estimate a slum index on Medellín, Colombia.

For other cities around the world, Stoler et. al. (2012)\cite{stoler2012assessing} and Weeks et.al (2007)\cite{stow2007object} used very high-resolution imagery for Accra (Ghana) deriving features based on texture and land cover to estimate a slums distribution index. Arribas-Bel, Patiño, and Duque (2017)\cite{arribas2017remote} employed a similar approach to estimate the Living Environment Deprivation index for small areas in Liverpool, England. Jean et. al. (2016)\cite{jean2016combining} used nighttime imagery with very high-resolution daily imagery to estimate variation in local-level economic outcomes for Nigeria, Tanzania, Uganda, Malawi, and Rwanda. They employed transfer learning of trained convolutional neural networks, a technique we applied in this case study as well.

\section{Methodology}
The open-source algorithm AP-LATAM, tries to reduce survey costs by analyzing high-resolution satellite imagery to detect potential areas of informal settlement growth. The final result is a geospatial dataset of areas that could contain informal settlements. By having updated potential slums growth locations, decision makers can prioritize the areas to survey. 

Source code for building and using a classifier to create datasets has also been released as open-source in a GitHub repository\footnote{\url{https://github.com/dymaxionlabs/ap-latam}} with a BSD-2 license. Instructions on how to train the classifier and predicting over new images are available there. This includes not only the steps for training the model but also the neural net weights for using with new satellite imagery. AP-LATAM is part of the Inter-American Development Bank's Code for Development platform\footnote{\url{https://code.iadb.org/en/repository/60/ap-latam}}.

This tool is already calibrated to detect slums and informal settlements in Buenos Aires, Asunción, and Montevideo. To obtain those results, Dymaxion Labs applied computer vision and deep learning algorithms to satellite imagery and other georeferenced data. Slum patterns such as texture and morphology of roofs are detected. From Dymaxion Labs experience calibrating these kinds of algorithms, such patterns have strong local dependencies. As TECHO has ground truth polygons and imagery from a previous survey in Guatemala City (Guatemala) in 2016, the strategy was based on training AP-LATAM on these data and then mapping potential informal settlements areas on Tegucigalpa (Honduras). Guatemala City is the capital and largest city of Guatemala, a country that borders with Honduras. It has 2,750,965 inhabitants with an area of 220 square kilometers. Taking into account the TECHO's survey database and the similarity of the geography of these cities, it was the best candidate to train AP-LATAM to map Tegucigalpa.

\begin{figure*}[t!]
    \centering
    \begin{subfigure}[t]{0.5\textwidth}
        \centering
        \includegraphics[scale=0.4]{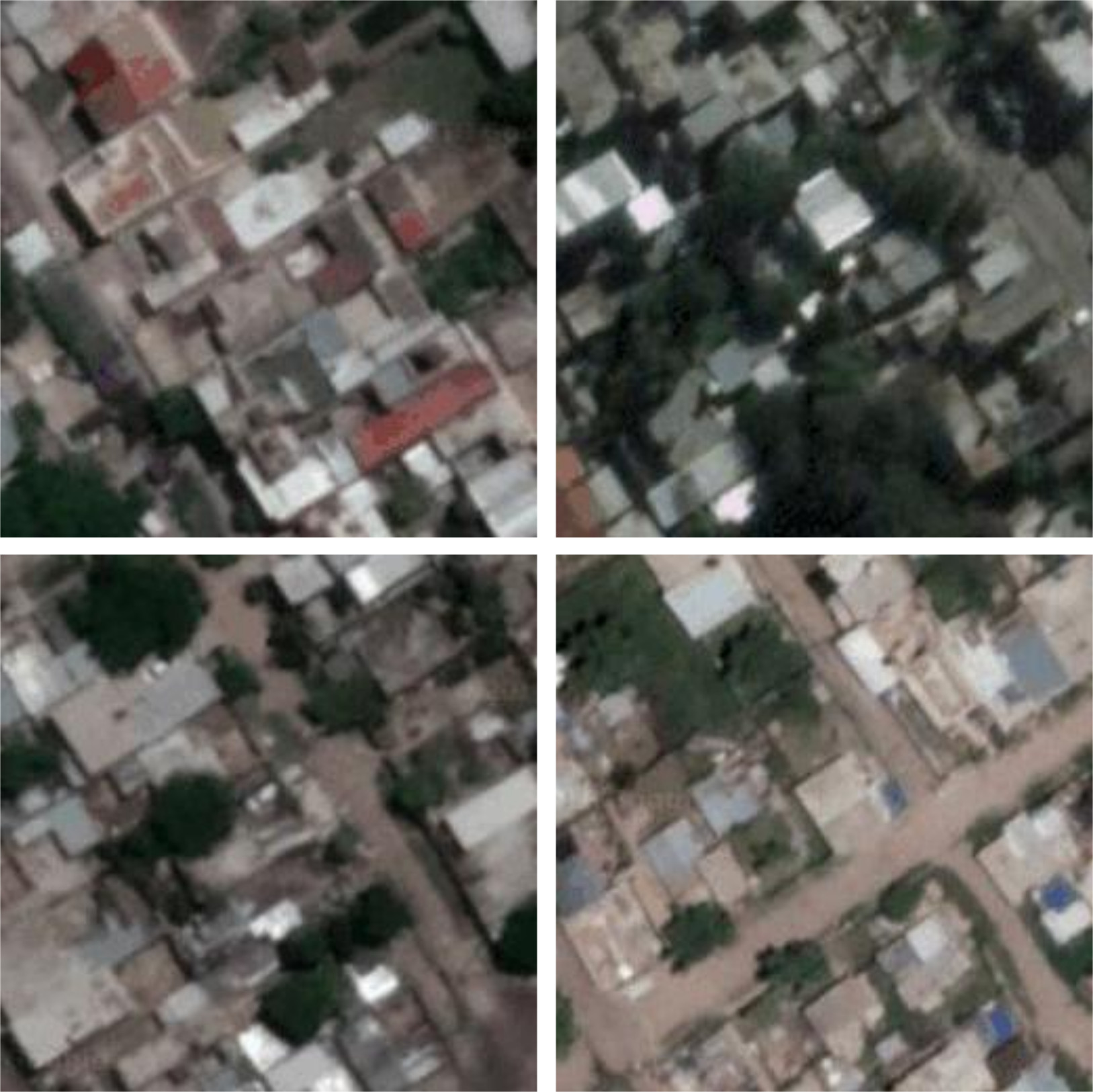}
        \caption{Tiles tagged as \emph{true}}
    \end{subfigure}%
    ~
    \begin{subfigure}[t]{0.5\textwidth}
        \centering
        \includegraphics[scale=0.4]{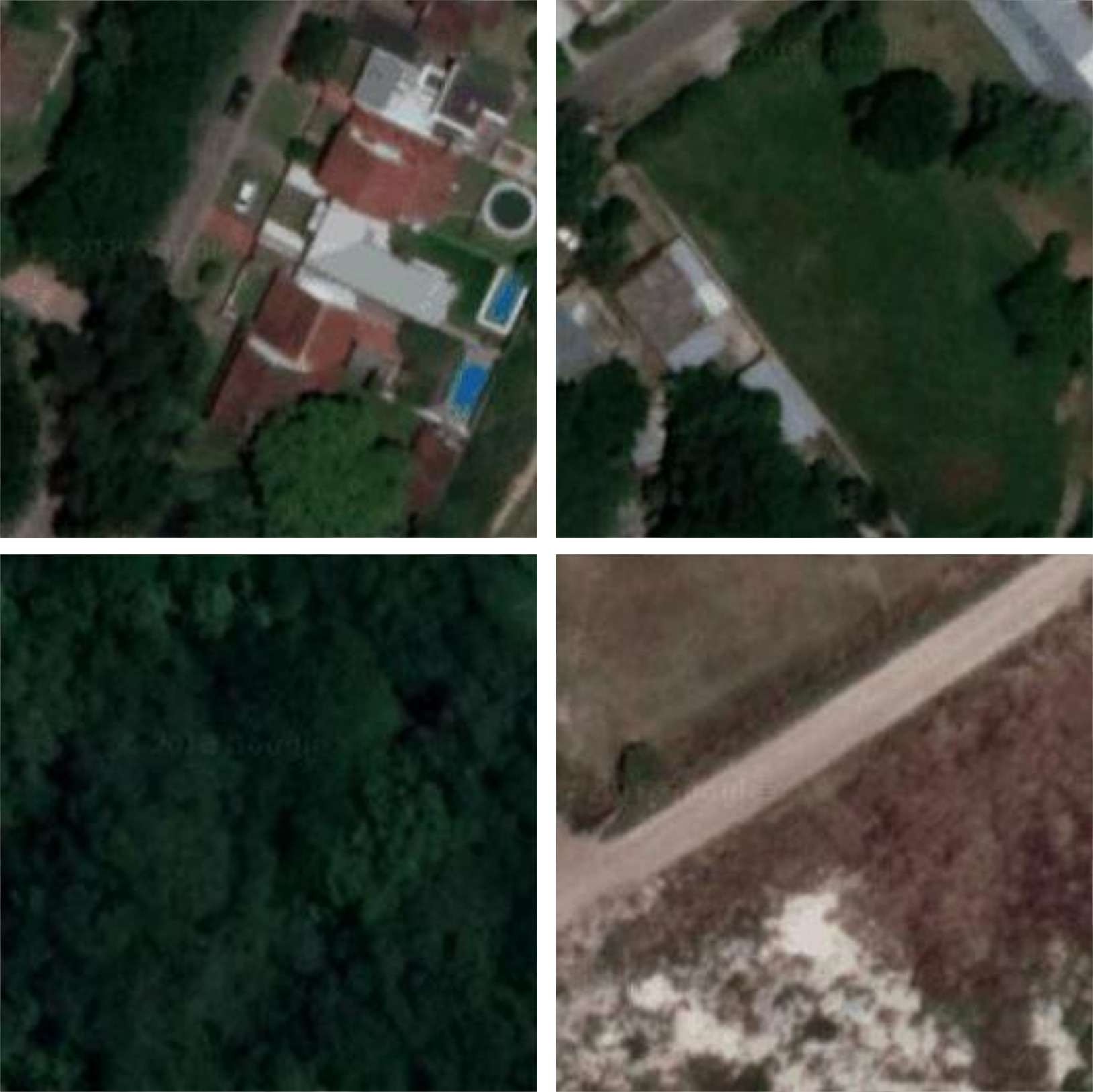}
        \caption{Tiles tagged as \emph{false}}
    \end{subfigure}
    \caption{Example of images tiles tagged as either true or false, whether they contain a informal settlement or not.}
    \label{fig:tiles}
\end{figure*}

The method consists of a binary classifier of image tiles of high-resolution sub-meter satellite imagery. Each image is classified as whether it contains an informal settlement or not. Some examples of these tiles area showed in figure \ref{fig:tiles}

To build the dataset for training and validation, the classifier takes a vector file of polygons of previously-known informal settlements and takes fixed-size tiles of images by sliding a window across the entire satellite image. For each image tile, it checks if the tile intersects with any polygon and tags it appropriately.

To make predictions over new images, it slides a window over the new image and builds a new vector file of polygons of the size of each positively-tagged image tile. The resulting dataset is post-processed to remove polygons with small probability and dissolve them into bigger polygons.

In our tests with Guatemala, as a final post-processing step, we used OpenStreetMap\cite{OpenStreetMap} datasets and crossed them with the polygons file.

\subsection{Imbalanced classes}
A binary classifier of informal settlements has imbalanced classes, that is, images tagged as positive (areas that contain settlements) are much less frequent than images tagged as negative (areas that do not contain settlements). To decrease bias and avoid overfitting, we under-sampled the negatives by taking a set of images of size proportional to the size of positives. We tried undersampling with a proportion of 4 and 8 and settled with 4.

\subsection{Data augmentation}
To help prevent overfitting and make the model generalize better, we perform data augmentation on the image tiles. We only applied horizontal and vertical flipping, but there are other random transformations that we could use to augment our dataset, like hue and brightness randomization (to account for differences in atmosphere corrections) and rotations.

\subsection{Fine-tuning}

A large amount of data is needed to build a functional convolutional neural network model. In practice, it is common to reuse a pre-trained network. However, most pre-trained networks work for a different set of labels and were not trained with satellite images, so for this use case it is necessary to retrain some of the top layers to improve prediction.

The methodology used here was to fine-tune a ResNet-50\cite{He2015} network with our satellite imagery. We chose this architecture taking into account the trade-off between precision and hardware requirements to perform experiments. The procedure is roughly as follows:

\begin{enumerate}
    \item Instantiate the convolutional base of ResNet-50.
    
    \item Add a fully-connected model on top, with a standard SGD optimizer and configure a binary cross-entropy loss function.
    
    \item Freeze the layers of the model up to the top 70 layers.
    
    \item Retrain the model.
\end{enumerate}

The Keras library was used for data augmentation, training, and prediction\cite{chollet2015keras}.

\subsection{Post-processing}

The resulting dataset after prediction over sliding windows is a set of small fixed-size squares, with a prediction probability associated. To refine the results we apply the following:

\begin{enumerate}
    \item \textbf{Median filter}: remove squares with low probability and a small number of neighbors.
    
    \item \textbf{Dissolve overlapping squares}: if the sliding window step size is smaller than the size of the windows, it may end up with overlapping squares, so this step dissolves them into a single polygon with mean probability values between the values of each connected squares.
\end{enumerate}

As mentioned before, we also used a dataset of blocks and calculated the intersection between the squares and blocks, and if sufficient squares covered a block, we picked them to form a new dataset of blocks that contain potential informal settlements. The outcome has better prediction accuracy mainly because roads and other areas are not considered.

\section{Results}

The resultant methodology had a Cohen's Kappa\cite{Cohen1960a} index of 0.82 at pixel level on Guatemala validation polygons. Based on the previous work done between both organizations analyzing the cities of Buenos Aires (Argentina), Asunción (Paraguay) and Montevideo (Uruguay), imagery from those countries were added to the training dataset. Due to the different geography, like the soil color in Asunción, and the common used materials for construction, the best results were achieved using only Guatemala data.

\begin{figure*}[h!]
 \includegraphics[scale=0.4]{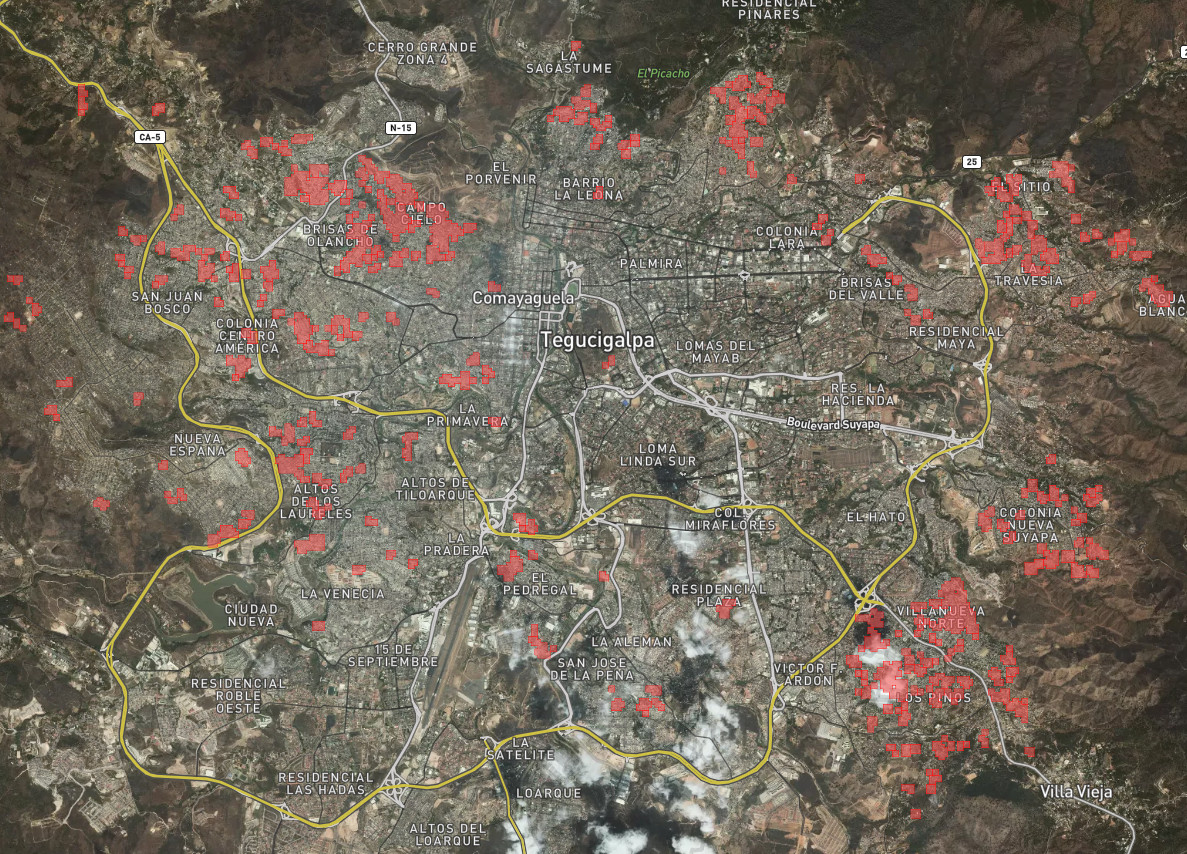}
 \caption{Tegucigalpa map of potential informal settlements.}
 \label{fig:map}
\end{figure*}

After being trained with the entire city, we applied it on the Tegucigalpa imagery to generate the map of potential areas for surveying. This map (figure \ref{fig:map}) was uploaded to our website (both as downloadable data and the online map). These datasets have been released as public domain data, using the same license\footnote{\url{https://opendatacommons.org/licenses/pddl/1-0/}} that OpenStreetMap uses for its data, the Open Data Commons Public Domain Dedication and License. From the AP-Latam website the user can download the datasets as GeoJSON files, one for each area and image acquisition date. The user can also explore an online map with the latest dataset generated.

Combining this map and other data sources from a utility company and government agencies, TECHO planned the fieldwork in two stages: territory exploration and surveyors training. To start the survey in one of the most violent cities in Latin America, a strict security protocol was designed. One of the protocols taken for the safety of the volunteers on the ground suggested not to survey an informal settlement without having previously contacted local referents that would allow them to guarantee minimum security. The search for local referents was carried out throughout the fieldwork. TECHO also used the potential informal settlements map to double-check the delimitation of the most hazardous places in the city to avoid sending volunteers to survey those areas.

\section{Conclusions}

The application of the transfer learning methodology trained in Guatemala City to map potential informal settlements in Tegucigalpa collaborates with TECHO's plan to survey the entire city. In particular, it helps them to map informal settlements located in dangerous areas where volunteers are not allowed to survey.

The 2018 Informal Settlements Census\cite{TECHOreporte} finally took place in the Metropolitan Area of the Central District (D.C.), specifically in the urban area, made up of the cities of Tegucigalpa and Comayaguela. The objective was to visit 796 neighborhoods and residential areas that make up the urban area, with the aim of identifying informal settlements and subsequently characterizing and georeferencing each one of them. The results were released as open data\footnote{\url{http://datos.techo.org/en/dataset/honduras-censo-de-asentamientos-informales-casco-urbano-distrito-central}}. 40\% of the territory where informal settlements without property titles are located belong to the state. Thereby, this map can be used by policymakers to relocate people living in risky areas immediately.

After the execution, 161 informal settlements were identified, 124 settlement files were raised with the help of references in informal settlements, and it was estimated that there are 42,000 families living in this situation.

From a replication perspective, having a technical team is critical not only for training and applying the algorithm, but also to help validate the mapping results. With more frequent ground validations, more iterations could be made on the algorithm to get a better accuracy in the final map. The technical skills and GIS knowledge of TECHO's team proved to be vital in achieving the quality level of ground surveys. In terms of mapping new cities, the availability of up-to-date high-resolution imagery is also essential to get relevant results.

\begin{acks}
To María Jesús Silva Rozas and Rubens C. Schrunder for giving us the chance to apply our methodology in a real-world case and validate the results in Honduras.

To UNICEF Innovation Fund team for their support as part of the Data Science cohort.
\end{acks}

\bibliographystyle{ACM-Reference-Format}
\bibliography{main}

\section{About the authors}

Federico Bayle is Bs. in Economics and MSc. in Data Mining and KDD. His master thesis was based on a machine learning methodology to map informal settlements in Buenos Aires, Argentina.

Damian Silvani is Bs. in Computer Sciences, with a specialization in Computer Vision and Machine Learning.
More than 15 years of experience in the software development industry solving hard problems.

\end{document}